# Sub-Attosecond Metrology via X-Ray Hong-Ou-Mandel Effect

S. Volkovich,[1] and S. Shwartz[1]

[1]Physics Department and Institute of Nanotechnology, Bar-Ilan University, Ramat Gan, 52900 Israel

We show that sub-attosecond delays and sub-Angstrom optical path differences can be measured by using Hong-Ou-Mandel interference measurements with x-rays. We propose to use a system comprising a source based on spontaneous parametric down-conversion for the generation of broadband x-ray photon pairs and a multilayer-based interferometer. The correlation time of the photon pairs and the Hong-Ou-Mandel dip are shorter than 1 attosecond, hence the precision of the measurements is expected to be better than 0.1 attosecond. We anticipate that the scheme we describe in this work will lead to the development of various techniques of quantum measurements with ultra-high precision at x-ray wavelengths.

Since its first observation [1], the Hong-Ou-Mandel (HOM) effect has attracted a great attention due to its importance for fundamental quantum sciences and since it holds a great promise for new quantum technologies [2-14]. The HOM effect is a quantum effect that is based on the interference of the wave functions of the photons rather than on the interference of classical waves. The striking consequence of this quantum interference is manifested when two indistinguishable photons arrive simultaneously at the two different input ports of a 50:50 beam splitter. In contrast to classical waves, the two photons will always be detected at the same output port of the beam splitter. As a result, coincidence measurements between the output ports are null as long as the photons at the two input ports are indistinguishable.

In a typical HOM experiment, two identical photons are generated and propagate along two paths. By varying one of the optical paths, it is possible to control the delay between the two photons so that they do not arrive at the beam splitter simultaneously and their distinguishability is raised. The more distinguished the photons become, the higher the probability of coincident detection gets. The ability of the HOM effect to detect the indistinguishability of photons on very short time scales has led to development of various approaches based on the effect for the measurements of ultrashort delays and optical path differences [15–18]. Measurements based on the HOM effect are more sustainable than measurements with classical interferometers, because unlike classical interferometers, HOM measurements are independent of the phase fluctuations of the optical beams. Consequently, in recent years several schemes for sub-femtosecond delay measurements with optical beams have been suggested and implemented [16,17].

Generally speaking, the extension of quantum optics to the x-ray regime can provide new intriguing opportunities. This is especially due to the availability of photon number resolving detectors demonstrating high detection efficiencies and negligible background noise. X-rays are also more penetrative than optical photons, and as they possess higher frequencies, they can be modulated to carry more information. We note that several works on quantum effects with x-rays have been reported recently [19]. For example, the necessity of a quantum theory to describe the effect of spontaneous parametric down conversion (SPDC) has been demonstrated [20], quantum effects such as electromagnetically induced transparency [21], collective Lamb shift [22], modulation of single γ photons [23], ghost imaging [24], quantum enhanced detection [25], and vacuum-assisted generation of atomic coherences [26] have been reported as well. In addition, several schemes for the generated x-ray polarization entangled photons have been proposed [27–29].

To reap the benefits of extending the HOM effect to the x-ray regime, a key requirement is a source that produces identical photons. One prominent candidate source is SPDC, where a pump photon interacts with the vacuum field in a nonlinear crystal to generate two photons (also known as biphotons) [30]. The two photons can be made indistinguishable by choosing the geometry and parameters of the system [1]. The keV bandwidth that has been reported for x-ray SPDC [20,31] suggests that the corresponding biphoton correlation time is on the order of a few attoseconds. The implementation of the x-ray HOM effect can lead to the development of quantum optical coherence tomography for measurements of very short spatial scales and tiny refractive index differences [32–35]. This would be appealing for the imaging of biological samples.

However, the possibility to measure such a broad spectrum HOM effect is not clear. The main challenge is that x-ray mirrors and beam splitters rely either on small angle reflection or on Bragg scattering [36]. Small angle reflection can be used to reflect a very broad spectrum, but the angular distribution of the generated photons is much broader than the acceptance angle of small angle reflection devices. Bragg scattering from crystals is narrow in both angle and spectrum, thus the HOM effect would be narrowband and the corresponding dip would be limited to a few femtoseconds. The alternative is to use Bragg scattering from artificial periodic structures made by multilayers. However, it is not clear whether the technical feasibility of the present-day technology allows the fabrication of such a system. It is also not clear a priori that the photons that hit upon the two input ports of the beam splitter are indeed indistinguishable.

In this letter we describe a system that is based on available technologies for measuring the HOM effect at x-ray wavelengths. We show that when the photons hit the beam splitter simultaneously, they are indeed indistinguishable, hence the system can support the detection of very short delays. We consider an example where the full width half max (FWHM) of the dip is about 0.6 attoseconds and explain how to control it.

We consider a standard scheme for HOM effect experiments, which consists a nonlinear crystal for the generation of x-ray biphotons, a phase shifter, two multilayer mirrors, and a multilayer beam splitter, as shown in Fig. 1. The biphotons emerging from the nonlinear crystal are commonly denoted as the "signal" and the "idler". This process is parametric, hence energy is conserved and $\hbar\omega_p = \hbar\omega_s + \hbar\omega_i$, where $\omega_p$, $\omega_s$, and $\omega_i$ are the angular frequencies of the pump, signal, and idler photons, respectively. The signal and idler propagate at different directions, which are symmetric with respect to the symmetry axis determined by the momentum conservation (phase matching) of the nonlinear process. Since all pertinent wavelengths are on the order of the distance between the atomic planes of the generating crystal, the phase matching is achieved by using the reciprocal lattice vector $\boldsymbol{G}$ [37]. This leads to the condition $\boldsymbol{k}_p + \boldsymbol{G} = \boldsymbol{k}_s + \boldsymbol{k}_i$, where $\boldsymbol{k}_p$, $\boldsymbol{k}_s$, and $\boldsymbol{k}_i$ are the wave vectors of the pump, signal, and idler, respectively. One of the biphotons is delayed by a phase shifter and each of them is reflected by a mirror and impinges upon the opposite side of a beam splitter.

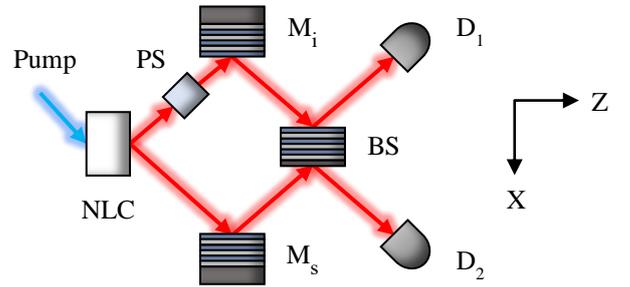

FIG. 1. Schematic diagram of the proposed experimental system. The pump photons are down-converted in the nonlinear crystal (NLC) into signal and idler photon pairs and the idler photon is delayed by a phase shifter (PS). The biphotons are then reflected by their corresponding multilayer mirrors ($M_i$ and $M_s$) into a beam splitter (BS) and the coincidence count rate at its output is measured by two detectors ($D_1$ and $D_2$).

The generation of the biphotons in the nonlinear crystal is described by the frequency domain coupled equations for the signal and idler envelope ladder operators in the Heisenberg picture for a lossless medium [20,31]. By using the

undepleted pump and the slowly varying envelope approximations we obtain

$$\frac{\partial \hat{a}_s}{\partial z} = \kappa \hat{a}_i^\dagger \exp(i\Delta k_z z),$$
$$\frac{\partial \hat{a}_i^\dagger}{\partial z} = \kappa^* \hat{a}_s \exp(-i\Delta k_z z). \quad (1)$$

Here $\hat{a}_s$ and $\hat{a}_i$ are the destruction operators of the signal and idler photons, respectively, $\kappa$ is a coupling coefficient, and $\Delta k_z = k_p cos\theta_p - k_s cos\theta_s - k_i cos\theta_i$ is the phase mismatch in the z direction. $\theta_p$, $\theta_s$, and $\theta_i$ are the angles between the lattice planes and the wave vectors of the pump, signal, and idler, respectively. The frequency domain operators are related to their real domain counterparts by $a_j(z, \boldsymbol{r}, t) = \iint_{-\infty}^{\infty} a_j(z, \boldsymbol{q}, \omega) \exp[-i(\boldsymbol{q} \cdot \boldsymbol{r} - \omega t)] d\boldsymbol{q} d\omega$, where $\boldsymbol{r} \equiv (x, y)$ and $\boldsymbol{q} \equiv (k_x, k_y)$, and they satisfy the commutation relations $[\hat{a}_j(z_1, \boldsymbol{q}_1, \omega_1), \hat{a}_k^\dagger(z_2, \boldsymbol{q}_2, \omega_2)] = \frac{1}{(2\pi)^3} \delta_{j,k} \delta(z_1 - z_2) \delta(\boldsymbol{q}_1 - \boldsymbol{q}_2) \delta(\omega_1 - \omega_2)$.

Assuming the low gain approximation, we solve Eq. 1 to obtain a transfer matrix for the SPDC source. To proceed it is more convenient to express the output of the SPDC crystal as a superposition of the vacuum state and the biphoton state, which is calculated from the transfer matrix

$$|\Psi\rangle = C|0\rangle + \int \int \int \int d\boldsymbol{q}_s d\omega_s d\boldsymbol{q}_i d\omega_i \\ \times \phi(\boldsymbol{q}_s, \omega_s, \boldsymbol{q}_i, \omega_i) \hat{a}_s^\dagger(\boldsymbol{q}_s, \omega_s) \hat{a}_i^\dagger(\boldsymbol{q}_i, \omega_i)|0\rangle. \quad (2)$$

Here $C$ and $\phi(\boldsymbol{q}_s, \omega_s, \boldsymbol{q}_i, \omega_i)$ are the probability amplitudes to detect the vacuum state and the frequency domain biphoton state, respectively. The obtained biphoton probability amplitude is

$$\phi(\boldsymbol{q}_s, \omega_s, \boldsymbol{q}_i, \omega_i) = (2\pi)^3 \kappa L e^{i\frac{\Delta k_z L}{2}} \operatorname{sinc}\left(\frac{\Delta k_z L}{2}\right) \\ \times \delta[\boldsymbol{q}_i - (\boldsymbol{q}_p + \boldsymbol{G} - \boldsymbol{q}_s)] \delta[\omega_i - (\omega_p - \omega_s)], \quad (3)$$

where $L$ is the crystal length.

Next, we consider the mirrors and the beam splitter. Multilayer devices are composed of alternating layers of two materials with low and high refractive indices, which are deposited on a substrate [36]. Since in the x-ray regime the refractive indices of materials depend only on the densities of the electrons, high and low atomic number materials are chosen to maximize the refractive index difference between the layers. These materials are commonly referred to as the "absorber" and "spacer", respectively, and their widths are indicated by $d_a$ and $d_s$. The ratio factor is defined as $\Gamma \equiv d_a/d$, where $d$ is the width of the bilayers.

We use Bragg's law with a correction for refraction [38] to find the necessary width of the bilayers for a specific wavelength and incidence angle. To estimate the required number of bilayers we use the recursive theory of multilayers [39]. It allows to obtain an analytical expression for the intensity reflectivity of $N$ bilayers, where the incident angle is equal to the Bragg angle and the refractions and the reflections from the substrate are negligible [38]

$$R = \tanh^2[2Nr\sin(\pi n\Gamma)]. \quad (4)$$

Here $r$ is the amplitude reflectivity of the electric field at the interface between the absorber and the spacer, and $n$ is the number of the Bragg peak. However, Eq. 4 provides only the reflectivity for a specific wavelength at the Bragg angle, while the down-converted photons contain many frequencies and angles. Thus we calculate numerically the reflectivity and the transmission of the mirror and the beam splitter by using the multilayer matrix theory [40], for the number of layers we estimated from Eq. 4, and obtain their transfer matrices.

Finally, we wish to calculate the count rate of coincidences between the two output ports of the beam splitter by using [41]

$$R_C = S \int \int d\boldsymbol{u} d\tau \\ \times \langle \Psi | \hat{a}_2^\dagger(\boldsymbol{r}_2, t_2) \hat{a}_1^\dagger(\boldsymbol{r}_1, t_1) \hat{a}_1(\boldsymbol{r}_1, t_1) \hat{a}_2(\boldsymbol{r}_2, t_2) | \Psi \rangle. \quad (5)$$

Here $S$ is the area of the pump beam at the input of the nonlinear crystal, $\boldsymbol{u} = \vec{r}_2 - \vec{r}_1$ is the distance between two detection points, $\tau = t_2 - t_1$ is the duration between the detections, and $\hat{a}_1$ and $\hat{a}_2$ are ladder operators at the two output ports of the beam splitter.

We calculate the propagation of the ladder operators through the system by using the transfer matrices of the multilayer devices and insert the result along with Eqs. 2 and 3 into Eq. 5. After a considerable but straightforward analytical calculation we obtain

$$R_C = \frac{S}{(2\pi)^9} \int \int d\boldsymbol{q} d\omega \{|M_s(\boldsymbol{q}_{++},\omega)M_i(\boldsymbol{q}_{--},\omega_p-\omega)\tilde{\phi}(\boldsymbol{q}_{++},\omega)|^2$$
$$\times [|A(\boldsymbol{q}_{+-},\omega_p-\omega)D(\boldsymbol{q}_{-+},\omega)|^2 + |B(\boldsymbol{q}_{-+},\omega)C(\boldsymbol{q}_{+-},\omega_p-\omega)|^2]$$
$$+M_s(\boldsymbol{q}_{+-},\omega_p-\omega)M_s^*(\boldsymbol{q}_{++},\omega)M_i(\boldsymbol{q}_{-+},\omega)M_i^*(\boldsymbol{q}_{--},\omega_p-\omega)\tilde{\phi}^*(\boldsymbol{q}_{++},\omega)\tilde{\phi}(\boldsymbol{q}_{+-},\omega_p-\omega)e^{i(\omega_p-2\omega)T}$$
$$\times [A(\boldsymbol{q}_{++},\omega)B^*(\boldsymbol{q}_{-+},\omega)C^*(\boldsymbol{q}_{+-},\omega_p-\omega)D(\boldsymbol{q}_{--},\omega_p-\omega)$$
$$+A^*(\boldsymbol{q}_{+-},\omega_p-\omega)B(\boldsymbol{q}_{--},\omega_p-\omega)C(\boldsymbol{q}_{++},\omega)D^*(\boldsymbol{q}_{-+},\omega)]\}. \quad (6)$$

Here we denote $\boldsymbol{q}_{\pm\pm} \equiv (\pm k_x, \pm k_y)$. The quantity $\tilde{\phi}(\boldsymbol{q},\omega) = \iint \phi(\boldsymbol{q}_s, \omega_s, \boldsymbol{q}_i, \omega_i) d\boldsymbol{q}_i d\omega_i$ is the biphoton probability amplitude, $T$ is the delay between the biphotons, $M_s$ and $M_i$ are the amplitude reflectivity of the signal and idler mirrors, respectively, and $A$, $B$, $C$, and $D$ are the elements of the transfer matrix of the beam splitter.

To demonstrate the feasibility to observe the effect, we consider an example of a system based on parameters that have been used in previous x-ray SPDC experiments [31]. The nonlinear crystal is a diamond crystal with a thickness of 0.8 mm and we use the C(660) atomic planes for phase matching. We assume that the pump is at 21 keV, its deviation angle from the Bragg angle is 8 mdeg and it is polarized in the scattering plane. The pump rate is $10^{13}$ photons/s and the area of the beam is 0.4 mm$^2$. The coupling coefficient in this case is estimated to have an order of magnitude of $10^{-19}$ m$^{-1}$ [31]. We choose the central photon energy of the signal and idler photons at 10.5 keV and the solution of the phase matching equation results in angles of propagation of 0.976 deg and -0.976 deg with respect to the optical axis described in Fig.1. The polarizations of the signal and idler photons are parallel, which is a result of this setup [31] and is required for identicalness. The corresponding Fresnel coefficients for both polarizations are approximately equal for our parameters.

We first show the spectrum of the coincidence count rate at the output of the nonlinear crystal in Fig. 2. We choose an aperture size of the detector of 0.4 deg, which defines the angular width of the SPDC and determines the photon energy range accepted by the detector to be 8.54 keV – 12.89 keV, due to the one-to-one correspondence between the energy and the propagation direction. The total rate is about 0.15 pairs/s and the bandwidth is 4.35 keV. This result agrees with the experimental results [31] and indicates on the possibility to measure delays with precision of sub-attosecond time scales.

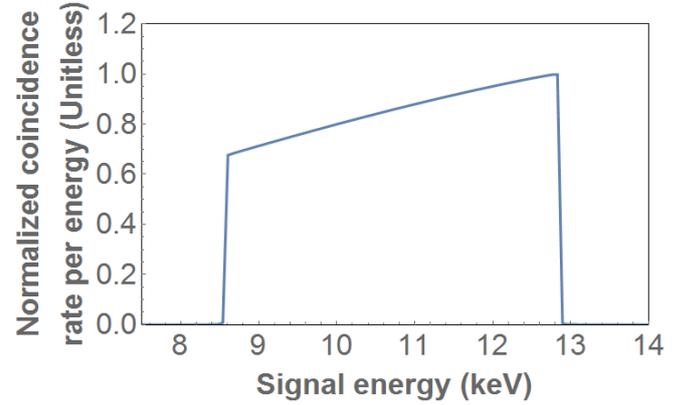

FIG. 2. The spectral dependence of the normalized coincidence count rate between the two output ports of the nonlinear crystal. The total bandwidth, which is obtained for a detector acceptance angle of 0.4 deg, is 4.35 keV.

Next, we consider the multilayer mirrors and beam splitter. Our goal is to show that it is possible to design optical devices with sufficient reflectivity that can accommodate the very broad angular distribution and spectrum of the generated biphotons. We choose the absorber layers to be platinum and the spacer layers to be carbon and we assume that the substrates are a silicon wafer. We use the data from [42] for the refractive indices and absorption coefficients. By using Eq. 4 we find that 20 bilayers with a width of 3.7 nm and with $\Gamma = 0.5$ are sufficient to achieve an intensity reflectivity of 90% and that 10 bilayers are sufficient to achieve approximately 50% reflectivity. For the beam splitter, the substrate width is 15 μm, which is shorter by an order of magnitude than the absorption length at 10.5 keV.

We simulate the dependence of the intensity reflectivity of the mirrors and the beam splitter on the incidence angle for

10.5 keV in Figs. 3(a) and 3(b). As expected, our simulation shows peaks in the reflectivity that obey Bragg's law. The high reflectivity at the lower angles is due to total reflection. We choose the first peak of the reflectivity at an incident angle of 0.976 deg, which is the incidence angle of the biphotons on the mirrors at perfect phase matching at the degenerate photon energy. The maximum of the reflectivity is 90% and the FWHM of the reflectivity of the mirror and the beam splitter are 0.07 deg and 0.095 deg, respectively. Figures 3(c) and 3(d) show the photon energy dependence of the reflectivity for an incident angle of 0.976 deg. The FWHM of the reflectivity of the mirror is 0.758 keV and of the beam splitter is 1.04 keV, whereas the bandwidth of the x-ray SPDC biphotons is 4.35 keV. Since the angular acceptance and the bandwidth of the multilayer devices are comparable to those of the biphotons, the parameters we select enable the observation of HOM dips at reasonable count rates.

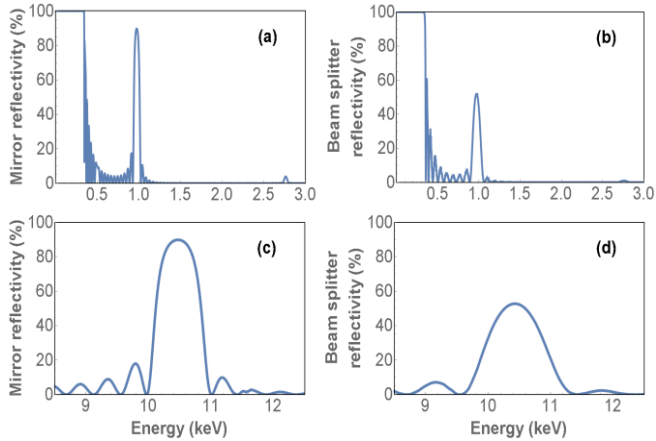

FIG. 3. The reflectivity of the multilayer mirror and the beam splitter as a function of the incidence angle, (a) and (b), and the photon energy, (c) and (d). Panels (a) and (c) show the mirror reflectivity and panels (b) and (d) the reflectivity of the beam splitter. The width of a bilayer is 3.7 nm, with $\Gamma = 0.5$.

Now we turn to the main result of this letter and show that the dip of x-ray HOM can be as short as 0.6 attoseconds at FWHM. We numerically calculate the integral described by Eq. 6 for various delays between the signal and the idler photons. Our results are shown in Fig. 4. The results are normalized to the output of the SPDC crystal, so they include the reflectivity, the transmission, the finite bandwidth, and the acceptance angles of the optical devices. It is clear that the dip of the coincidence count rate is nearly zero. The FWHM of the dip indicates on a correlation time of about 0.6 attoseconds, which corresponds to a spectral bandwidth of 1.097 keV. This ultrashort time scale corresponds to an optical path difference between the two arms of the HOM setup of about 1.8 Angstroms.

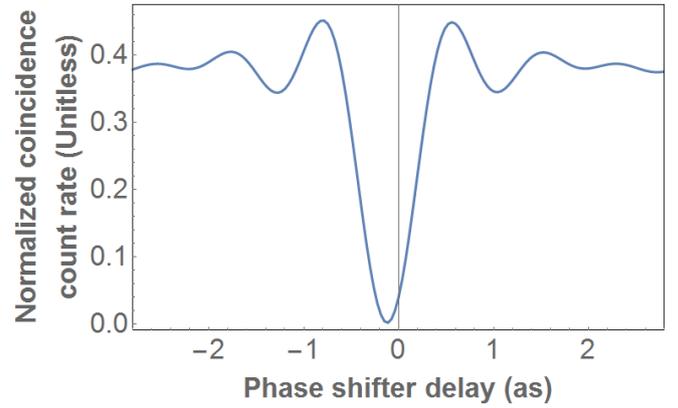

FIG. 4. The normalized coincidence count rate between the two output ports of the beam splitter as a function of the delay between the biphotons. The width of the predicted dip is about 0.6 attoseconds at FWHM. The shift from zero is due to the slight difference in the paths of the biphotons. See text for details.

We note that the energy bandwidth we calculated is wider than the bandwidth in Figs. 3(c) and 3(d). However, this is not surprising since those figures show the bandwidth for a specific incident angle, while the angular distribution of the biphotons is broad. This observation indicates on the possibility to observe even shorter dips by designing multilayer devices with an angular dispersion that matches that of the biphotons. Also, we note that when the biphotons impinge on the beam splitter, one of them propagates through the substrate first. This asymmetry causes a small phase differences between the amplitude reflectivity of the two beam splitter ports, which leads to a shift in the coincidence count rate dip. It does not, however, destroy their indistinguishability, since the intensity reflectivity is nearly equal for both sides.

We emphasize that we have described here an example for possible parameters. However, our simulations show that the x-ray HOM effect can be measured for a large range of parameters. We stress that the stability with respect to mechanical vibrations can be improved by using narrower band optical devices or narrower detector apertures, but in the cost of widening the dip in the coincidence count rate. This may be overcome by using advanced data analysis procedures [17,18]. Alternatively, fabrication of the system as a monolithic structure would improve the stability significantly.

We also emphasize that while short time delays and optical path differences can be measured with x-ray interferometers [43–45], the HOM system exhibits several important advantages. Since in the HOM effect the interference is between the wave functions of the biphotons and not between classical coherent beams, the experiment can be performed by using incoherent sources, whereas standard interferometers require sources with high spatial coherence. Another advantage is the requirements for stability of the effect, which are less stringent than the requirements for interferometers. While interferometers have to be more stable than the wavelength for the entire measurement, thus on the angstrom scale for x-rays, in the HOM effect the system has to only be stable enough to maintain the biphotons indistinguishable during a detection cycle.

In conclusion, we have described how to implement the Hong-Ou-Mandel effect in the x-ray regime and how to utilize the effect for the measurement of sub-attosecond time intervals and sub-angstrom optical path differences. The relaxed requirements for stability and for the coherence of the source in comparison to interferometers suggest that the effect can be used for a large class of measurements for fundamental science and for a variety of applications. We note that the approach we describe can be performed with present day x-ray sources although the expected count rate is quite moderate. New advanced sources such as the new high repetition rate free-electron lasers [46,47] are expected to enhance the count rate significantly. Consequently, our work opens the possibility for quantum precision measurements that are supported by the ultra-high spatio-temporal precision that is enabled by using quantum effects with x-rays.